\documentclass[letterpaper,twocolumn,10pt]{article}
\usepackage{usenix2019_v3}

\usepackage{caption}
\usepackage{graphicx,color,listings}
\usepackage{flushend}
\usepackage{tabularx}
\usepackage{authblk}

\def\System{Venkman}
\def\Comment#1{}

\title{Restricting Control Flow During Speculative Execution with {\System}}

\author{Zhuojia Shen}
\author{Jie Zhou}
\author{Divya Ojha}
\author{John Criswell}
\affil{Department of Computer Science\\
University of Rochester}

\begin{document}
%-------------------------------------------------------------------------------

    \maketitle

    \begin{abstract}
Side-channel attacks such as Spectre that utilize speculative execution
to steal application secrets pose a significant threat to modern computing
systems.  While program transformations can mitigate some
Spectre attacks, more advanced attacks can divert
control flow speculatively to bypass these protective
instructions, rendering existing defenses useless.

In this paper, we present \emph{\System}: a system that employs
program transformation to completely thwart Spectre
attacks that poison entries in the Branch Target Buffer (BTB) and the
Return Stack Buffer (RSB).  {\System} transforms code so that all valid
targets of a control-flow transfer have an identical alignment in
the virtual address space; it further transforms all branches to
ensure that all entries added to the BTB and RSB
are properly aligned.  By transforming all code this way,
{\System} ensures that, in any program wanting Spectre defenses, all
control-flow transfers, including speculative ones, do not skip over
protective instructions {\System} adds to the code segment to mitigate
Spectre attacks.  Unlike existing
defenses, {\System} does not reduce sharing of the BTB
and RSB and does not flush these
structures, allowing safe sharing and reuse among programs while
maintaining strong protection against Spectre attacks.
We built a prototype of {\System} on an IBM POWER8 machine.  Our evaluation
on the SPEC benchmarks and selected applications
shows that {\System} increases execution time
to 3.47$\times$ on average and increases code size to 1.94$\times$ on average
when it is used to ensure that fences are executed to
mitigate Spectre attacks.  Our evaluation also shows that
Spectre-resistant Software Fault Isolation (SFI) built using {\System} incurs
a geometric mean of 2.42$\times$ space overhead and
1.68$\times$ performance overhead.
\end{abstract}

    %%=============================================================================
\section{Introduction}
\label{section:intro}
%%=============================================================================

%
% Outline:
%
% o Speculation attacks pose significant threat
% o Some defenses exist for Spectre Variant-1
% o Spectre Variant-2 poses a significant challenge
%   o Poisons BTB and RSB
%   o Undermines all compiler-based defenses
% o Limitations for defenses for Spectre Variant-2
%   o Retpolines do not work
%   o Intel features have limitations
%     o Fails to protect code running at same hardware privilege level
%     o Reduces utilization of BTB
%     o Reduces performance
%     o Only works if processor provider adds support
% o Our method
%   o It works
%   o It does not reduce speculation
%   o Does not limit sharing of BTB
%

Spectre attacks~\cite{Spectre:Oakland19} pose a significant threat to
computing systems.  Such attacks can be launched by unprivileged code
and leverage speculative execution within
processors to trick victim programs into leaking confidential data.
Current Spectre attacks first direct control flow into infeasible program
paths which load sensitive data (often loaded from an out-of-bounds
array access) into a processor register and then cause the victim to
leak the information via a side channel.  To date, Spectre has been
used by malicious processes to steal information from other victim
processes~\cite{Spectre:Oakland19}, by malicious JavaScript code to
steal information from the web browser~\cite{Spectre:Oakland19}, and by
malicious code to steal secrets contained within Trusted
Execution Environments (TEEs) like Intel SGX~\cite{Foreshadow:UsenixSec18}.
Spectre attacks work on Intel and AMD x86 processors and on
ARM processors~\cite{Spectre:Oakland19}, and this work
demonstrates that Spectre attacks work on the IBM POWER8 processor.
Any processor implementing speculative execution and branch
prediction is likely vulnerable to Spectre attacks.
Consequently, Spectre poses a significant threat to nearly
every laptop, desktop, and server computer.

Variant-1 Spectre attacks~\cite{Spectre:Oakland19}
require that victim programs contain specific code patterns that are
exploitable during speculative execution.  Variant-2 Spectre attacks which
poison the processor's Branch Target Buffer (BTB)~\cite{Spectre:Oakland19}
or Spectre variants that poison the
Return Stack Buffer (RSB)~\cite{SpectreRSB:WOOT18,Ret2Spec:CCS18}
direct a victim's control flow to small pieces of code that exhibit the same
behavior even if such paths are infeasible in the program's
non-speculative control flow.
Spectre attacks which poison the
BTB or RSB~\cite{Spectre:Oakland19,SpectreRSB:WOOT18,Ret2Spec:CCS18} are
especially
nefarious as attackers can use them to bypass instructions inserted into the
program that mitigate Spectre attacks.  For example, Intel suggests
adding fence instructions to mitigate
Spectre Variant-1~\cite{IntelSpecMitigate}, and
Dong et al.~\cite{SpectreVG:HASP18} propose a software fault isolation
(SFI)~\cite{SFI:SOSP93} mechanism that operates correctly even when
subjected to a Spectre Variant-1 attack.
However, if an attacker poisons the BTB or RSB, speculative execution
can jump straight to load instructions without first executing the
fence or SFI instructions.

For some processors, there are microcode updates that allow the
operating system (OS) kernel to limit sharing of the BTB and provide
new mechanisms for flushing the BTB and RSB~\cite{IntelSpecMitigate}.
However, only processors that use microcode can be updated without a
physical processor replacement, and microcode updates are not
available for all afflicted processors~\cite{IntelMicroCode:2018}.
The only software defense is the retpoline; versions exist for mitigating
BTB poisoning~\cite{Retpoline} and
RSB poisoning~\cite{Ret2Spec:CCS18}.
However, all retpolines contain an unconditional direct branch
instruction.  Evidence shows that direct branches consult the BTB just
as computed branches do~\cite{JumpOverASLR:MICRO16}.  This means that
retpolines are likely vulnerable to BTB poisoning.  Additionally,
retpolines use return instructions to branch to target addresses.
This makes them inherently incompatible with control-flow integrity
(CFI) defenses~\cite{CFI:TISSEC09}.

% Unfortunately, existing defenses against BTB and RSB poisoning are
% incomplete.  Instructions that restrict the sharing of BTB
% entries or flush the BTB~\cite{IntelSpecMitigate} can lead to
% suboptimal performance.  New hardware instructions only mitigate
% attacks between software at different privilege
% levels~\cite{IntelSpecMitigate}.  To the best of our knowledge, all
% existing defenses mitigate attacks against the BTB but not the
% RSB~\cite{IntelSpecMitigate,Retpoline}, allowing poisoning of the
% RSB~\cite{SpectreRSB:WOOT18,Ret2Spec:CCS18}.

We present a new, comprehensive, software-only defense against Spectre
attacks that poison the BTB and RSB.  Named \emph{\System}, our solution
performs two transformations on code to
mitigate Spectre attacks.  The first transforms \emph{all} code on a system
so that it cannot poison the BTB and RSB with target addresses of the
attacker's choosing. {\System} transforms programs to group
instructions into \emph{bundles}.  All bundles have the same
power-of-two length and are aligned at the same power-of-two boundary
in memory.  {\System} then transforms the program so that all computed
branches first align the target to a bundle boundary.  The net effect
is that all BTB and RSB entries for all code on the system are addresses at
the beginning of a bundle.  We also propose a system
architecture that ensures that all binary code running on a system has
been transformed as described, ensuring that only aligned bundle addresses are
inserted into the BTB and RSB.
The second transformation modifies
programs wanting defense against Spectre attacks by adding
instructions into their bundles that mitigate Spectre attacks.
For example, this second transformation can insert fence instructions
into each bundle to mitigate all Spectre attacks (as suggested by
previous defenses for Spectre variant-1~\cite{IntelSpecMitigate}), or
it can insert Spectre-resistant SFI instructions~\cite{SpectreVG:HASP18}
into bundles containing load instructions.
Since {\System} ensures that instructions
that need protection from Spectre attacks (such as loads) are in the
same bundle as the instructions providing the protection (such as
fences), training of the BTB and RSB cannot cause execution to bypass
the instructions providing protection.
The second transformation can be excluded on programs that don't want
Spectre defenses.

%\Comment{JTC: We need to add text below that shows that the POWER8
%architecture is an important architecture.  Examples of important
%real-world systems using POWER8 would be very helpful.}

%\Comment{JTC: ZS, please update overhead summaries to
%account for real-world applications results.}

We built a prototype of {\System} for the POWER architecture.  Our
IBM POWER8 machine employs speculative and out-of-order execution, and we
have demonstrated that the Spectre proof-of-concept
code~\cite{Spectre:Oakland19}, ported to POWER and changed to poison the
BTB, works on our
POWER8 machine.  We evaluated the performance
of our prototype on the SPEC CPU 2017 benchmarks and on several
real-world applications.  In geometric means,
our results show that the bundling transformation incurs a
performance overhead of 1.09$\times$ and a space overhead of 1.61$\times$.
Our results also show
that {\System}, when used to ensure that fence instructions are
executed to mitigate Spectre attacks, increases
execution time to 3.47$\times$ on average and
increases code size to an average of 1.94$\times$.
We have also used {\System} to build a Spectre-resistant sandbox using
SFI; this system has a geometric mean of 2.42$\times$ space
overhead and 1.68$\times$ performance overhead.

To summarize, our contributions are as follows:

\begin{itemize}
  \item
  We have designed a complete software-only solution that
  prevents poisoning of the BTB and RSB.
  To the best of our knowledge, {\System} is the first complete
  software-only solution to such attacks.

  \item
  We have evaluated the overheads that {\System}'s padding and alignment
  transformations incur and found that our solution induces a geometric
  mean of 1.61$\times$ space overhead and 1.09$\times$ performance overhead.
  %We
  %believe these overheads are acceptable for preventing software from
  %poisoning BTB and RSB entries.

  \item
  We have evaluated the performance overheads of adding barrier instructions
  to mitigate Spectre attacks in code requiring defense from Spectre
  attacks.  We found that our solution induces a geometric
  mean of 1.94$\times$ space overhead and 3.47$\times$ performance
  overhead.

  \item
  We have evaluated {\System} in providing an SFI system that
  resists Spectre attacks with a geometric mean of 2.42$\times$ space
  overhead and 1.68$\times$ performance overhead.
\end{itemize}

The rest of the paper is organized as follows.
Section~\ref{section:spectre} provides background on Spectre attacks.
Section~\ref{section:model} describes our threat model.
Section~\ref{section:design} describes the design of our defense, and
Section~\ref{section:impl} describes the implementation of our prototype.
Section~\ref{section:security} conducts an empirical study of {\System}'s security guarantees.
Section~\ref{section:results} presents our space and performance evaluation of {\System},
Section~\ref{section:related} examines related work, and
Section~\ref{section:conc} concludes and discusses future work.

    %%=============================================================================
\section{Background on Spectre Attacks}
\label{section:spectre}
%%=============================================================================

% What is Spectre?
Spectre
attacks~\cite{Spectre:Oakland19,SpectreRSB:WOOT18,Ret2Spec:CCS18,Spectre1.1}
are a family of attacks that leverage speculative execution to trick a 
victim program into speculatively executing a sequence of instructions which 
it normally would not execute. Although the processor will eventually
revert the architectural effects of speculatively executed
instructions~\cite{Shen:MPD:1},
the execution of the instructions may change the state of internal
structures (such as the caches and branch prediction buffers) within the
processor.
A typical Spectre attack consists of two basic steps. First, a processor is 
tricked into speculatively executing instructions chosen by an adversary
which load secret data into the processor's registers. Second, 
the adversary uses a side channel, such as
a {\sc Flush+Reload}~\cite{FlushReload:UsenixSec14} attack on the caches,
as a covert channel to leak the
secret information in the registers to the attacker.
There are two major variants of Spectre attacks. 
One exploits conditional branches; the other poisons 
indirect branches and
returns~\cite{Spectre:Oakland19,SpectreRSB:WOOT18,Ret2Spec:CCS18}. 

%%=============================================================================
\subsection{Exploiting Conditional Branches}
\label{section:spectre:conditional}
%%=============================================================================

% \Comment{JTC: JZ, I think this section needs to clearly explain the
% code patterns that Spectre Variant 1 utilizies to leak data.  You will
% then need to state in the next subsection that Spectre Variant 2 looks
% for ``gadgets'' that it can execute speculatively in the case that the
% program doesn't have the patterns it needs for Variant 1.}

\begin{figure}[tb]
\begin{lstlisting}[language={C},caption={Conditional Branch Example},label={fig:branch}]
  if (x < arr1_boundary) {
      y = arr2[arr1[x] * 256];
  }
\end{lstlisting}
\end{figure}

For a conditional branch, modern processors predict whether the branch
will or will not be taken and
proceed to speculatively execute the instructions that it predicts are needed
next.
This keeps the processor pipeline busy, increasing throughput~\cite{Shen:MPD:1}.

Consider the conditional branch code in Listing~\ref{fig:branch}.
During a Variant-1 Spectre attack~\cite{Spectre:Oakland19},
before the
condition
at Line~1 is resolved, the processor proceeds to speculatively execute code
at Line~2 if the branch predictor predicts that the condition is true.
After the processor determines that {\tt x} is not less than the
length of the array {\tt arr1},
it reverts changes it has made to registers.  However, data brought
into the cache during speculative execution remains in the cache.
Therefore, {\tt arr2}'s element whose index is {\tt arr1[x] * 256}
is still within the cache even though it is not needed.
If the variable {\tt x} is controlled by an attacker, and a secret value is 
located at {\tt arr1[x]},
then the attacker can infer the secret data by
using a side-channel attack such as
{\sc Flush+Reload}~\cite{FlushReload:UsenixSec14}
or {\sc Prime+Probe}~\cite{PrimeProbe:CT-RSA06}.

Intel~\cite{IntelSpecMitigate} recommends placing a load fence
({\tt lfence}) instruction~\cite{IntelArchManual16}
before instructions reading memory that are control-dependent on
branches; a fence makes sure that all prior instructions are retired
before executing subsequent instructions, ensuring that the load
executes only if it was supposed to be executed.

%%=============================================================================
\subsection{BTB and RSB Poisoning}
\label{section:spectre:indirect}
%%=============================================================================

\begin{figure}[tb]
\begin{lstlisting}[language={C},caption={Indirect Branch Example},label={fig:indirect-branch}]
  (*func_ptr)();

  if (x < arr1_boundary) {
      load_fence();
      y = arr2[arr1[x] * 256];
  }
\end{lstlisting}
\end{figure}

For an indirect branch, before the destination address is resolved, the
processor consults the BTB (or RSB if the indirect branch is a return
instruction) to predict the next address from which to fetch
instructions~\cite{Shen:MPD:1}. Similar
to branch prediction, this optimization improves the processor's throughput.

However, an adversary process can poison the BTB and RSB and trick a victim 
process into speculatively executing code gadgets chosen by the
adversary~\cite{Spectre:Oakland19,SpectreRSB:WOOT18,Ret2Spec:CCS18}.
Two features enable BTB and RSB poisoning.
First, all processes running on the same physical CPU core share the 
same BTB and RSB~\cite{Ge:JCE16}. Second, the operating
system kernel does not save or flush the state of the BTB and RSB when
context switching between processes or threads, allowing one process
to add values into the BTB and RSB that are used by a subsequent
process running on the same core.
Therefore, a malicious process can mistrain the BTB and
RSB to fill in target addresses to which it wants a victim to jump.
For example, in Listing~\ref{fig:indirect-branch}, an attacker could train the
BTB entry for the call through a function pointer at Line~1 so that it
speculatively jumps straight to Line~5, bypassing the load fence placed
before the load at Line~4.  With BTB and RSB poisoning, an attacker can
trick a victim
process into executing any code within the victim's code segment; the
attacker can target code that leaks sensitive information as described
in Section~\ref{section:spectre:conditional}.

While current Spectre attacks use BTB poisoning to alter speculative
control flow for indirect branches~\cite{Spectre:Oakland19}, we
believe they can also change the speculative control flow of direct branches.
Evtyushkin et al.~\cite{JumpOverASLR:MICRO16} demonstrated that an
unconditional jump can be used to create BTB collisions; they trained
the BTB entries used by direct branches to launch side-channel attacks
against the OS kernel.
Even though the target of a direct branch is specified as an immediate operand
to the instruction, the processor may still use the BTB to predict the
target of a direct branch; this allows the processor to start fetching
instructions at the target of the branch during the fetch stage of the
processor pipeline before learning that an instruction is a branch during
the decode phase of the pipeline~\cite{Shen:MPD:1}.

The poisoning of targets for direct branches threatens to break
retpolines.  Retpolines~\cite{Retpoline,Ret2Spec:CCS18} are
the only existing software defense for Spectre attacks that poison the
BTB and RSB.  Retpolines use direct
branches to jump to code within the retpoline.  An attacker could
poison the BTB entry used by the direct branch, causing the retpoline
to speculatively execute code at an attacker's desired location.
While existing Spectre attacks have only exploited the indirect
branches~\cite{Spectre:Oakland19}, we believe that direct branches are
likely to be exploited at some point.  Unlike
retpolines~\cite{Retpoline,Ret2Spec:CCS18},
{\System} can mitigate any type of BTB poisoning as it ensures that only
valid addresses are inserted into the BTB and RSB.

%Intel uses BTB for conditional
%and unconditional jumps~\cite{JumpOverASLR:MICRO16} whereas Power8 processors
%uses branch
%prediction for every conditional branch but not for unconditional
%branches~\cite{powermanual}.

%%=============================================================================
\subsection{Read-Only Protection Bypass}
\label{section:spectre:codecorruption}
%%=============================================================================

Similar to how Meltdown~\cite{Meltdown:UsenixSec18} leverages the late
enforcement of user/supervisor protection flags, an attacker can
exploit the late checking of read and write permissions on pages to
change the value read from a write-protected memory location during
speculative execution~\cite{Spectre1.1}.  Speculative stores combined
with store-to-load forwarding could corrupt the code segment if the
instruction fetch unit can read values speculatively written to memory
that are stored in the processor's store buffer.  This could allow an
attacker to modify instructions added by a compiler that mitigate
Spectre attacks.  For example, the attacker could replace fences or
SFI instructions with NOP instructions, effectively disabling the
protections.  Any defense using compiler instrumentation techniques
must defend against speculative modifications to the code segment.

    %%=============================================================================
\section{Threat Model}
\label{section:model}
%%=============================================================================

%
% Outline:
%
% o Any piece of code can be an attacker
% o Only a subset of programs are potential victims
% o Hardware is implemented correctly according to ISA
%

Our threat model assumes that an attacker will attempt to steal data
with a Spectre attack~\cite{Spectre:Oakland19} that poisons the
BTB and/or RSB.  Since some processors predict direct branch
targets using the BTB~\cite{JumpOverASLR:MICRO16}, we assume that the
attack can poison both direct and indirect branches and calls.
Our model is restricted to attacks that
leak data through the cache via data accesses.  Other potential attacks
that leak data through other side channels, such as the instruction cache,
translation look-aside buffers (TLBs),
branch predictors, and functional units, are out of scope.

Our model is very broad: any piece of software
may be a potential attacker.  This includes all user-space software as well
as the OS kernel.  A subset of software on the system comprises
potential victims of the attack.  This model covers attacks by one
application against another, attacks by a distrusted part against
another within one application, attacks launched by applications against
the OS kernel, and attacks launched by a compromised OS kernel against
applications (similar to the Foreshadow attack~\cite{Foreshadow:UsenixSec18}).

Our model assumes that the hardware is implemented correctly with
respect to the processor's instruction set architecture (ISA), meaning
that the processor updates memory and processor registers correctly
but that speculative execution performed by the processor may allow
Spectre attacks~\cite{Spectre:Oakland19} to leak information through
microarchitectural state.

    %%=============================================================================
\section{Design}
\label{section:design}
%%=============================================================================

Spectre variants that poison the BTB~\cite{Spectre:Oakland19} and
RSB~\cite{SpectreRSB:WOOT18,Ret2Spec:CCS18} work because one
program can insert entries into the BTB and RSB that are correct for its
address space but incorrect for a victim program in the victim program's
address space.  {\System} transforms all code running on the system so
that \emph{any} BTB or RSB entry created by one program does not cause any
other program on the system to bypass fence instructions (or other
instructions inserted by a compiler) that protect load instructions from
Spectre attacks.
{\System} transforms code so that instructions are grouped into bundles
and then instruments branches to ensure that they can only target the
first instruction in each bundle.  As long as load instructions and the
instructions that protect them are within the same bundle, attackers
cannot execute a load without first executing the protecting
instructions.  In short,
{\System} ensures that branches can only insert the initial address of
a bundle into the BTB and RSB.

\emph{All} code running on a system must be transformed as described above.
We first describe a
software architecture that can ensure that \emph{all} code on a
system has been transformed using {\System}.
We then present how {\System} lays out the virtual address space of
the system to facilitate its instrumentation and
how {\System} transforms code to ensure that only ``safe''
code addresses are inserted into the BTB and RSB.
Finally, we present how
{\System} can be used to ensure that fences are used to prevent loads
from accessing invalid memory and how {\System} can be integrated with
Spectre-resistant SFI~\cite{SpectreVG:HASP18} to provide
speculation-safe sandboxing.

%%-----------------------------------------------------------------------------
\subsection{{\System} Architecture}
\label{section:design:arch}
%%-----------------------------------------------------------------------------

\begin{figure}[t]
  \centering
  \resizebox{\columnwidth}{!}{%
    \includegraphics{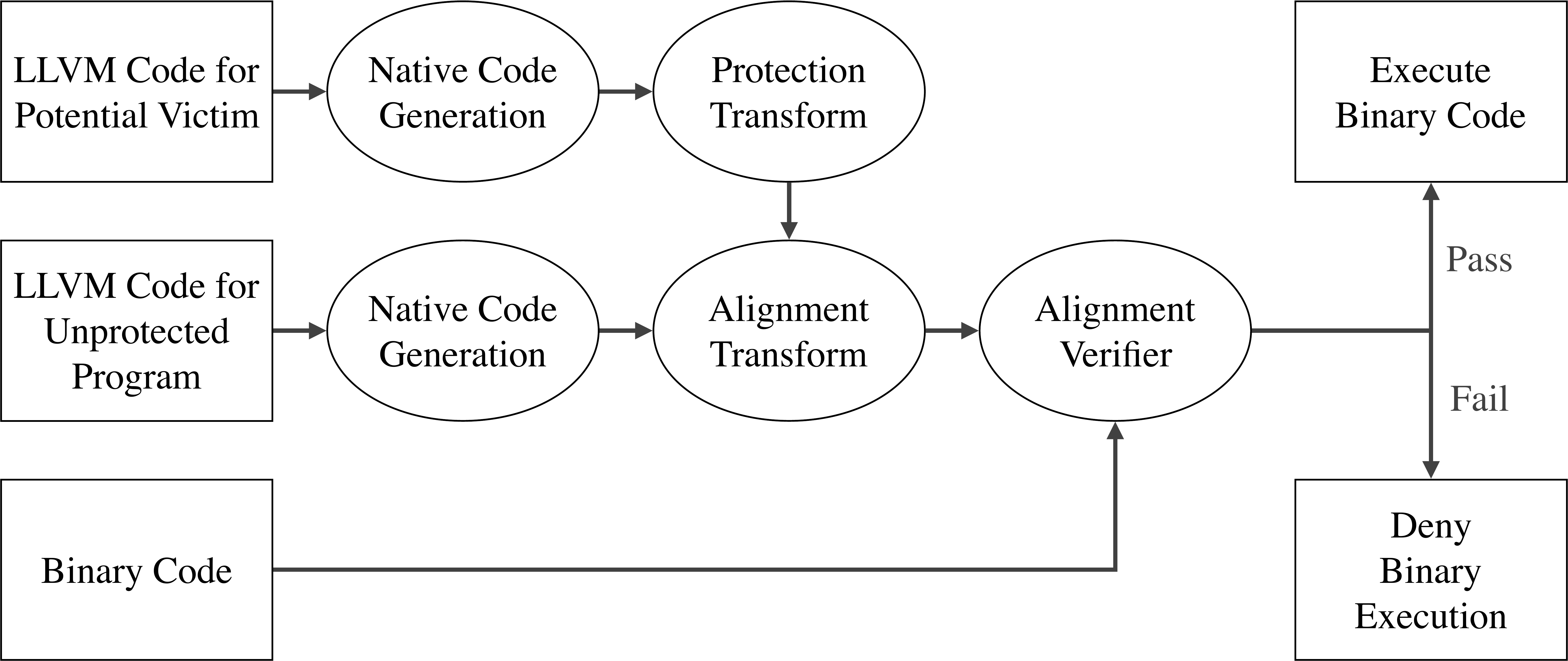}
  }
  \caption{{\System} Architecture}
  \label{fig:arch}
\end{figure}

{\System} must ensure that all code running on the system is transformed so
that the code only inserts ``safe'' values into the BTB and RSB
as Sections~\ref{section:design:align} and~\ref{section:design:cfi}
will describe.  We therefore need a system that can transform code on site
if possible and verify that binary code from third parties has already
been transformed by {\System}.

Figure~\ref{fig:arch} shows {\System}'s architecture.  {\System}
supports programs encoded in one of two formats.  The first format is
native binary code; this is how software is shipped today.  {\System}
cannot statically transform such programs due to the challenges of
static binary rewriting e.g., accurate disassembly of the native code
and reconstruction of its control-flow graph (CFG).  However, a
binary verifier can verify that the native code within the executable
has already been transformed as {\System} requires.  Binaries can, for
example, come with
Typed Assembly Language (TAL) annotations~\cite{TAL:Morrisett:TOPLAS99}
which can help the verifier efficiently prove that the native code conforms to
{\System}'s requirements.  Since the compiler that created the binary
must have transformed it, it can easily insert the TAL annotations
required for verification.  A verifier like RockSalt~\cite{RockSalt:PLDI12}
could be modified to perform the verification that {\System} requires.

The second program format is LLVM Bitcode~\cite{LLVM:CGO04,LLVM:MICRO03}.
This format represents programs in a virtual instruction set that
makes program analysis and transformation efficient and accurate.
The LLVM virtual instruction set~\cite{LLVM:CGO04} organizes a program
as a set of functions; each function has an explicit CFG,
alleviating the need to reconstruct the CFG from
binary code.  Instructions in LLVM virtual instruction set are in
Static Single Assignment (SSA) form~\cite{TheSSAPaper},
allowing efficient SSA-based algorithms to be employed to analyze code.
An extended version of the LLVM virtual instruction set
can encode a complete OS kernel completely within the LLVM instruction
set~\cite{SVA:SOSP07}, so both application code and OS kernel code
can be shipped in LLVM Bitcode format.

Once {\System} generates native code for an LLVM Bitcode executable,
it passes the native code (annotated with CFG information)
through an optional set of
transformations that add instructions to mitigate Spectre attacks
e.g., fence instructions as Intel recommends~\cite{IntelSpecMitigate}.
The code is then passed through the alignment transformations
(described in Sections~\ref{section:design:align} and~\ref{section:design:cfi})
that prevent the program from poisoning BTB and RSB entries.  Once
transformed, the verifier checks that the BTB and RSB defenses
have been applied correctly before allowing the code to execute.  By
reusing the binary verifier to verify the native code it generates,
{\System} removes its compiler transformations and native code
generator from its Trusted Computing Base (TCB).

On a system running {\System},
the OS kernel and dynamic binary loader must already
have been transformed with {\System}.  Additionally, the OS
kernel must ensure that programs do not modify or extend their
code segments without the binary verifier first verifying the new
code.  This can be accomplished by modifying the
{\tt mmap()} and {\tt mprotect()} system calls in the
OS kernel so that they verify code within a page before making the
page executable.
Systems such as SecVisor~\cite{SecVisor:SOSP07} can ensure
that all kernel code has been verified before it is loaded.

% \paragraph{Virtual Instruction Set Computing} A second method is to
% integrate {\System} into Secure Virtual Architecture
% (SVA)~\cite{SVA:SOSP07,SVAOS:UsenixSec09}.  SVA is a compiler-based
% virtual machine situated between the software stack and the processor;
% it translates virtual instruction set code to native code and,
% consequently, can control native code generation of all code (both
% user-space and kernel-space) on the system.  SVA also provides code
% segment integrity, ensuring that software cannot execute native code
% that SVA has not itself translated from virtual instruction set
% code~\cite{SVAOS:UsenixSec09}.  {\System} could be integrated into
% SVA's native code generator, allowing it to be combined with SVA-based
% systems that enforce memory safety~\cite{SVA:SOSP07,SVAOS:UsenixSec09}
% or control-flow integrity~\cite{KCoFI:Oakland14} or that protect
% applications from a compromised OS
% kernel~\cite{VirtualGhost:ASPLOS14,Apparition:UsenixSec18}.

%%-----------------------------------------------------------------------------
\subsection{Virtual Address Space Layout}
\label{section:design:layout}
%%-----------------------------------------------------------------------------

\begin{figure*}[tb]
\centering
\resizebox{\textwidth}{!}{%
  \includegraphics{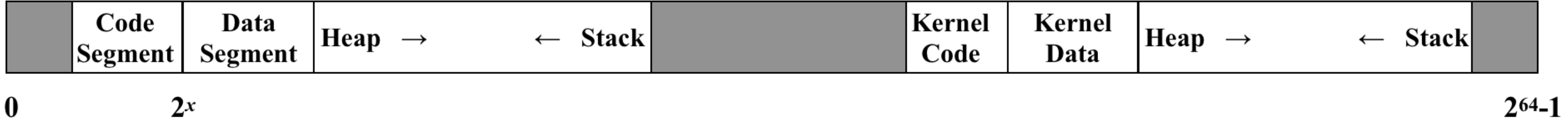}
}
\caption{{\System} Virtual Address Space Layout}
\label{fig:vmlayout}
\end{figure*}

{\System} divides the virtual address space as Figure~\ref{fig:vmlayout}
depicts.  {\System} places the application code and data in the lower portion
of the virtual address space and kernel code and data in the upper
portion of the virtual address space; this arrangement is used by many
current operating systems, including
Linux~\cite{LinuxKernel3:BovetCesati} and
FreeBSD~\cite{FreeBSDKernelV2}.

All application code must be located within the lower contiguous
portion of the virtual address space denoted as \emph{code segment} in
Figure~\ref{fig:vmlayout}.  Unlike existing systems, this code segment
includes code
loaded by the dynamic linker.  Likewise, all kernel code, including the code
for dynamically-loaded kernel modules, must be loaded within the region
reserved for kernel code shown in Figure~\ref{fig:vmlayout}.
{\System} must ensure that all entries in the BTB and RSB are
addresses within the application code segment or the kernel code
segment.
By requiring that all code be loaded within the application or kernel
code segment, and by strategically selecting the placement and size of
these code segments, simple bit-masking of control data can easily
ensure that all computed branches target an address in the application
code segment (for application code) or the kernel code segment (for
kernel code). 

We evenly split the code area and the data area (Data Segment, Heap, and Stack 
in Figure~\ref{fig:vmlayout}) in the user space; with this split,
{\System} can identify to which area an address belongs by checking the
highest bit (which is the {\it x}'th bit in Figure~\ref{fig:vmlayout}).
This helps
us implement SFI~\cite{SFI:SOSP93} more easily.
Section~\ref{section:impl} explains this choice in more detail.

Our implementation, described in Section~\ref{section:impl}, reserves
32~KB of virtual address space at the beginning and at the end of the
code segment; Figure~\ref{fig:vmlayout} denotes these
reserved areas with dark gray boxes. 
This platform-dependent change allows {\System} to enforce SFI with lower 
overhead on a POWER machine.  Section~\ref{section:impl} explains the
reasons in detail.

%%-----------------------------------------------------------------------------
\subsection{Code Alignment}
\label{section:design:align}
%%-----------------------------------------------------------------------------

\begin{figure}[tb]
\centering
\resizebox{\columnwidth}{!}{%
  \includegraphics{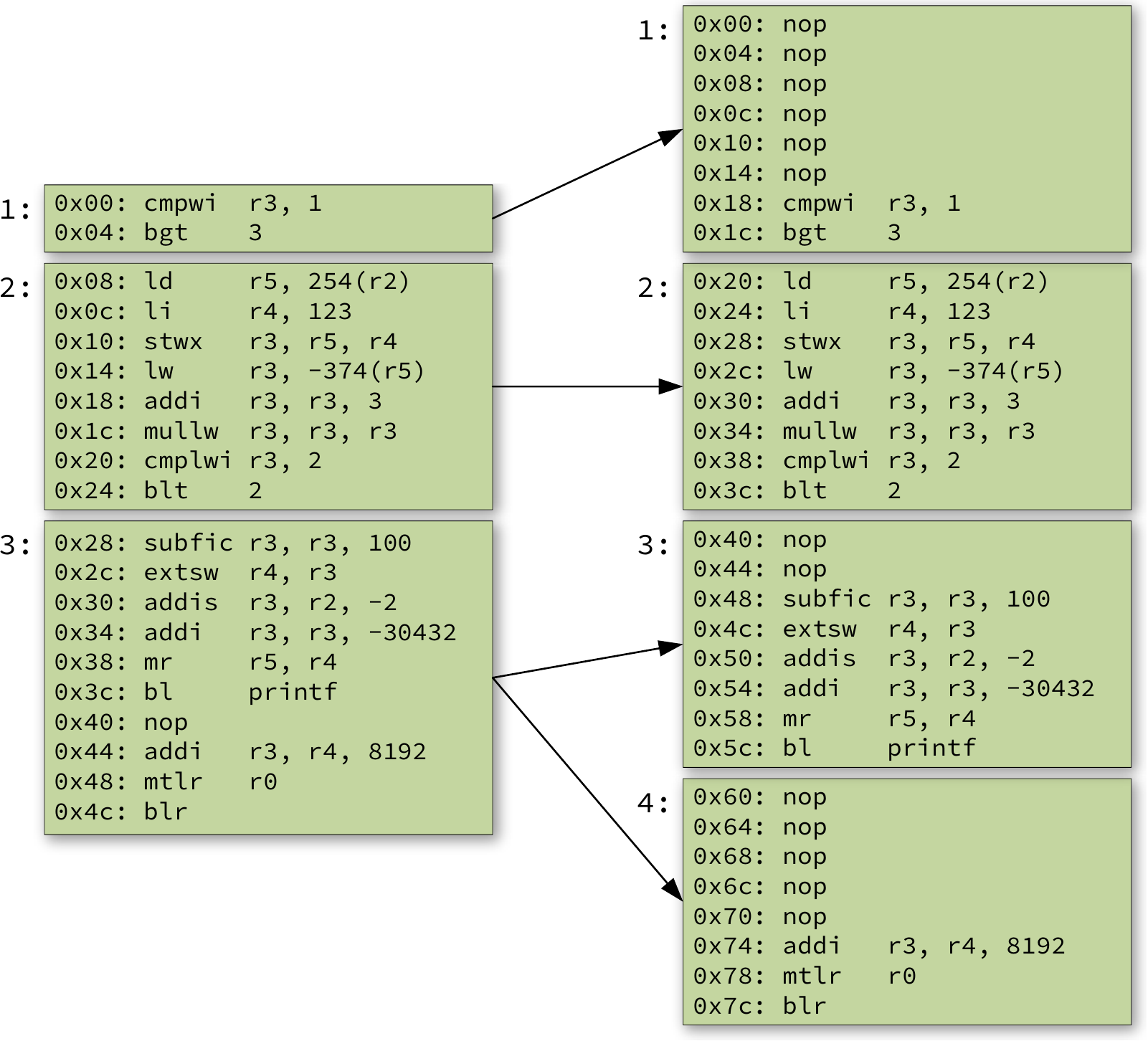}
}
\caption{Example of {\System} Code Alignment with 32-Byte Bundles}
\label{fig:bundle}
\end{figure}

{\System} transforms code so that each basic block within the program
has a power-of-two size and is aligned at an address divisible
by the same power-of-two.  For example, {\System} can divide a
program's basic blocks into bundles of 32 bytes (8 instructions per bundle on a
POWER machine)
with each bundle aligned on
a 32-byte boundary as Figure~\ref{fig:bundle} shows.  Basic
blocks larger than the required size are broken into smaller basic
blocks, and basic blocks smaller than the required size are padded
with NOPs until they are the required size.  This simple
transformation creates an invariant for all targets of control flow
transfers: each address to which a branch or call instruction can jump
is aligned on a specific boundary (a 32-byte boundary in our example).
{\System} ensures that all entries in the BTB and RSB fulfill this
invariant with static and dynamic checks;
Section~\ref{section:design:cfi} explains how.

When transforming basic blocks to have the correct size and alignment,
{\System} must enforce several restrictions.  First, any instruction
needing protection from a Spectre attack must appear in the same basic
block as the instructions that protect it.  For example, if a defense
against Spectre inserts a load fence before a load instruction, the
load fence and the load must occur within the same bundle.  Second,
targets of control flow transfers must be at the start of a basic
block.  The alignment of basic blocks ensures that the targets
of branch and call instructions are the first address of a bundle.
However, return addresses require additional processing: call
instructions must always occur at the end of a bundle so that the
return address is the beginning of the next bundle in memory.

%%-----------------------------------------------------------------------------
\subsection{Control Flow Instrumentation}
\label{section:design:cfi}
%%-----------------------------------------------------------------------------

With all basic blocks properly sized and aligned, {\System} must
enforce the following two invariants on entries in the BTB and RSB:

\begin{enumerate} \item{\textbf{Code Segment}:} All entries in the BTB
and RSB must be within the application code segment or kernel code
segment.  If the processor checks page permissions late in the
pipeline or uses a unified TLB, failure to ensure this invariant may
cause the processor to speculatively execute data as code.

	\item{\textbf{Alignment}:} All entries in the BTB and RSB must be
	aligned to the first address of a basic block (in our running
	example, a 32-byte boundary bundle).  This prevents the program from
	bypassing instructions such as fences during speculative execution.
	\end{enumerate}

Furthermore, {\System} must enforce these invariants even if the
program has memory safety errors.

To do this, {\System} must ensure that all branches, jumps, and calls
target only the first address of an aligned basic block at run-time.
{\System} must enforce this requirement on direct jumps and calls,
indirect jumps and calls, and return addresses.  The processor may use
the BTB on both direct and indirect jumps to determine the next
address from which to fetch instructions, and it will similarly use
the RSB for return instructions~\cite{Shen:MPD:1}.

Direct control flow transfers are correct by construction: the target
address is always the beginning of a specific basic block.  Since
{\System} aligns the basic block, the target of direct jumps to a
basic block are aligned by construction.

For indirect control flow transfers such as calls through registers
and returns, {\System} must insert bit-masking instructions before the
indirect jump, indirect call, or return instruction that align the
target of the branch to the beginning of a bundle. 
For example, before the return instruction {\tt blr} in the last bundle in
Figure~\ref{fig:bundle}, {\System} inserts instrumentation instructions to
ensure that the link register (which contains the return address) always point
to the beginning of a bundle.
Additionally, the bit-masking must ensure that the target address is within the
program's code segment.  By placing the code segment strategically in
memory, simple bit-masking of the higher-order bits suffices.

On processors supporting target operands in memory (e.g., {\tt ret}
instructions on x86~\cite{IntelArchManual16}), {\System} must
transform the code so that the target address is first loaded into a
register, bit-masked as described above, and then used in an indirect
branch instruction that takes its argument in a
register.  Otherwise, one thread could corrupt the address in memory
and change its alignment after the bit-masking has occurred but before
the indirect jump uses the target in memory.

The above changes ensure that all values in the BTB are aligned to a
basic block entry point within the code segment.  For the RSB, since
call instructions are placed at the end of a basic block, the return
address is constructed to be at the beginning of the next aligned
basic block.  This ensures that all RSB entries are the address of an
aligned basic block.

%%-----------------------------------------------------------------------------
\subsection{Speculative Stores to the Code Segment}
\label{section:design:sfi-store}
%%-----------------------------------------------------------------------------

SFI~\cite{SFI:SOSP93} can be used to
prevent an application from speculatively reading and writing memory
regions to which it does not have access~\cite{SpectreVG:HASP18}.  On
processors that can forward the results of speculative stores to instruction
fetches (as Section~\ref{section:spectre:codecorruption} describes),
we can use Spectre-resistant SFI~\cite{SpectreVG:HASP18} to ensure
that the instructions that {\System} adds do not get speculatively
overwritten by speculative store instructions.  This SFI
instrumentation inserts instructions before
each store instruction to ensure that the address used in the store
instruction is outside of the application and kernel code segments.

%The idea here is to assign code and data to be in separate memory
%locations using linker directives and checking the stores not to be
%storing to code locations at the runtime.

%%-----------------------------------------------------------------------------
\subsection{{\System} with Other Defenses}
\label{section:design:fence}
%%-----------------------------------------------------------------------------

{\System} provides a framework into which other defenses can be integrated.
For example, existing Spectre attacks~\cite{Spectre:Oakland19}
leak secrets through the data cache by using a load on an address that
is computed from the secret data.  {\System} can ensure
that all instructions retire before each load to prevent
leaks through the data cache that do not exist within the program's
non-speculative control flow.
For each bundle that has at least one load instruction, {\System}
can insert a
barrier instruction, such as x86's {\tt lfence}~\cite{IntelArchManual16}
or POWER's {\tt eieio}~\cite{Power8Manual},
at the beginning of the bundle to guarantee that all the instructions
executed prior to loads
in the bundle are executed and retire first.
With {\System}, this approach completely thwarts Spectre
Variant-1, Variant-2, and other variants that poison the RSB.
{\System} can also use Spectre-resistant
SFI techniques~\cite{SpectreVG:HASP18} to prevent load instructions
from accessing a specific region of the virtual address space; this
creates a speculation-resistant sandbox for the application.
Section~\ref{section:impl} explains in more detail how we integrate
other defenses into {\System}.

    %%=============================================================================
\section{Implementation}
\label{section:impl}
%%=============================================================================

%%-----------------------------------------------------------------------------
\subsection{Base {\System} Implementation}
\label{section:impl:cfi}
%%-----------------------------------------------------------------------------

We implemented {\System} by extending the code generator in the
LLVM~\cite{LLVM:CGO04} 4.0 compiler with two new MachineFunction Passes.
To create a prototype of {\System} as quickly as possible, we
opted to build our prototype for the POWER architecture first.
POWER remains an important and competent architecture for high
performance computing, cloud computing, and enterprise-level
workloads~\cite{PerfOptIBMPower}.  Our IBM POWER8
machine utilizes speculative out-of-order execution and is therefore
vulnerable to Spectre attacks.  Beneficially, POWER has
fixed-sized instructions which hastened development of our
prototype~\cite{Power8Manual}.
We leave x86 and ARM implementations of {\System} for future work.

\begin{figure}[tb]
\begin{lstlisting}[caption={CFI Instrumentation},label={fig:cfi}]
  clrrdi r1, r1, 5
  clrldi r1, r1, 19
  mtlr   r1
  blr
\end{lstlisting}
\end{figure}

The first MachineFunctionPass inserts the code that bit-masks indirect
branch targets as Section~\ref{section:design:cfi} describes.
On POWER, indirect branch targets are stored either
in the link register or the counter register~\cite{Power8Manual}.  The
first MachineFunctionPass searches for instructions that move values
into the link register and counter register and inserts instructions
to clear the lower-order 5 bits (because we choose to use 32-byte bundle). 
Since the counter register is also
used for purposes other than indirect branch targets, our
MachineFunctionPass scans for the next use of the counter register and
only bit-masks the value if the next use of the counter register
within the same function is a branch instruction.   

Listing~\ref{fig:cfi} shows an example for a return address (stored
in register {\tt r1} that is moved into the link register with the {\tt mtlr}
instruction.  The {\tt clrrdi} instruction clears the lower 5 bits of the
code pointer stored in {\tt r1} so that it is aligned on a bundle
boundary.  The {\tt clrldi} instruction clears the upper 19 bits of
the code pointer to ensure that it is located within the lower 32~TB of
the virtual address space where we put the code segment, as
Section~\ref{section:impl:sfi} describes.

The second MachineFunctionPass breaks up basic blocks and aligns them
as Section~\ref{section:design:align} describes.  It ensures that all
indirect control flow instructions and the bit-masking inserted by the
previous MachineFunctionPass remain within the same bundle.
Our implementation uses 32-byte bundle, so they are aligned on 32-byte
boundaries.

%%-----------------------------------------------------------------------------
\subsection{{\System} with Fences}
\label{section:impl:eieio}
%%-----------------------------------------------------------------------------

Our prototype has a command-line option to insert a
barrier instruction ({\tt eieio}) in each bundle containing at least
one load instruction.  The {\tt eieio} instruction enforces ordering
of memory accesses issued prior to the barrier with memory accesses
issued after the barrier~\cite{Power8Manual}.
By placing an {\tt eieio} in each bundle
that contains one or more load instructions, we can mitigate Spectre
attacks completely by ensuring that all checks on the pointer used in load
instructions retire before the load instruction commences.
We therefore use our {\tt eieio} option
in experiments to evaluate the performance of {\System} when it is
used to ensure that {\tt eieio} instructions are executed before load
instructions to mitigate Spectre attacks.

%%-----------------------------------------------------------------------------
\subsection{{\System} with SFI for Stores and Loads}
\label{section:impl:sfi}
%%-----------------------------------------------------------------------------

\begin{figure}[tb]
\begin{lstlisting}[caption={SFI Instrumentation on Stores},label={fig:sfi-store}]
  rldicr r1, r1, 32, 63
  ori    r1, r1, 0x2000
  rldicl r1, r1, 32, 18
  std    r3, 8(r1)
\end{lstlisting}
\end{figure}

\begin{figure}[tb]
\begin{lstlisting}[caption={SFI Instrumentation on Loads},label={fig:sfi-load}]
  clrldi r1, r1, 1
  ld     r3, 8(r1)
\end{lstlisting}
\end{figure}

We also implement SFI for {\System}.  Our prototype provides SFI on stores
and (optionally) SFI on loads.  SFI on stores prevents store-bypass
attacks~\cite{Spectre1.1} from speculatively modifying the code segment.
SFI on both stores and
loads can provide Spectre-resistant isolation for software plugins.
We therefore implemented sandboxing using Spectre-resistant
SFI~\cite{SpectreVG:HASP18} for the POWER architecture.

In order to achieve an efficient SFI implementation, we divide the whole
virtual address space for user programs ({\tt 0} to
{\tt 0x3fffffffffff}) equally into two regions, one ({\tt 0} to
{\tt 0x1fffffffffff}) for the code segment and the other
({\tt 0x200000000000} to {\tt 0x3fffffffffff}) for data segments
(which includes global variables, thread stacks, and the heap).
As Section~\ref{section:design:layout} describes, in this way we could
easily identify to which region a pointer is pointing by examining the
highest bit used in addressing user space memory (which is bit 45).
Furthermore, since POWER's D-Form store instructions~\cite{Power8Manual}
allow the
target address to be the sum of a register operand's contents and a
16-bit signed immediate, {\System} must reserve a 32-KB hole between {\tt 0}
and the beginning of the code segment and another 32-KB hole between the
end of the code segment and the beginning of the first data segment.
In this way, neither a store with a register operand pointing to the
beginning of the data segment and a negative immediate, nor a
store with a register operand pointing to the end of virtual address
space and a positive immediate, can speculatively overwrite the
code segment.  This narrows down the code segment region
to {\tt 0x8000} to {\tt 0x1fffffff7fff}.  With this arrangement,
before every store
instruction, our prototype inserts code that clears bit 45 of the
pointer register content used by the store to ensure that it points
outside of the
virtual address region reserved for the code segment.

To evaluate the overhead of SFI on loads for programs that want to
employ sandboxing, we implemented an optional feature to add SFI
instrumentation on loads.  Since
there is no special memory region needing protection from
speculative loads in our testing programs, we choose to instrument every load
to ensure that each reads user-space (as opposed to kernel-space)
memory; this is done by inserting code that clears the most significant bit of
the pointer.  This instrumentation is mainly for the purpose
of mimicking the code size and performance overhead of incorporating
SFI on loads into our defenses; it also happens to
prevent Meltdown~\cite{Meltdown:UsenixSec18} attacks on the OS kernel.

Listing~\ref{fig:sfi-store} shows the instrumented code for a D-Form
store.  The {\tt std} instruction takes the sum of the contents of
register {\tt r1} and an immediate 8 as the target address, and it
stores the contents of register {\tt r3} as a double word into the target
address.  Our instrumentation rotates the contents of the pointer by 32
bits, sets bit 13 (bit 45 of the actual pointer), rotates the pointer
back, and clears bits 46 to 63.
This ensures that the pointer always points to memory in the data
segments and prevents speculative writes to the code segment.

Listing~\ref{fig:sfi-load} shows the instrumentation for a D-Form load.
The {\tt clrldi} instruction clears the most significant bit of the
contents of register {\tt r1}, which is used in the {\tt ld} instruction
as a pointer from which to load.  The instrumentation prevents kernel space
memory from being speculatively read by a user space application.

On POWER, load and store instructions have another form called
X-Form~\cite{Power8Manual}.  X-Form loads and stores compute the
target address by adding the contents of two registers (a base and an
index register).  For such loads and stores, our prototype adds code to
add the base and index register (placing the result in the base
register), bit-mask the result, and use the result in a D-Form
instruction that replaces the original X-Form instruction.
A subsequent subtract instruction restores the original contents of the
base register.

Our prototype does not use the POWER predicated move instruction,
{\tt isel}, because our MachineFunction Passes execute after register
allocation.  Consequently, use of compare and {\tt isel} instructions
can overwrite the condition registers.  Our current instrumentation
overwrites no condition registers and can be safely inserted anywhere
before a store or a load.

%%-----------------------------------------------------------------------------
\subsection{Limitations}
\label{section:impl:limit}
%%-----------------------------------------------------------------------------

Our current implementation has two limitations.
First, we didn't instrument the OS kernel with {\System}.
Second, we didn't instrument most of the C/C++ standard library code;
the GNU C Library (glibc) and C++ Library (libstdc++) are tightly bound
with the GNU C/C++ compiler (gcc and g++) and thus cannot be compiled by
LLVM/Clang entirely.  Instead, we opted to compile most of glibc and
libstdc++ code by gcc and g++ with alignment options\footnotemark and
instrument only {\tt exit.c},
{\tt msort.c}, {\tt elf-init.c}, and {\tt libc-start.c} in glibc by
{\System} separately.  These files contain initialization routines and
library functions that call functions in application code.  Since
{\System} inserts bit-masking instructions that clear the last few bits
of indirect branch targets, an instrumented callee in application code
might return to an incorrect address in its unaligned caller in library
code.  For a similar reason, we also didn't bit-mask indirect branch
targets in some application functions including {\tt main} and all
global constructors and destructors of C++ programs.  These functions
are called by library code that is not instrumented by {\System}.

\footnotetext{{\tt -falign-functions=32}, {\tt -falign-jumps=32},
{\tt -falign-labels=32}, and {\tt -falign-loops=32}.}

    %%=============================================================================
\section{Security Evaluation}
\label{section:security}
%%=============================================================================

To evaluate {\System}'s effectiveness,
we implemented a proof-of-concept Spectre Variant-2
attack on an IBM POWER8 machine and
demonstrated that {\System} with fences successfully prevents the attack.

\begin{figure}[tb]
\begin{lstlisting}[language={C},caption={Core Component of Attack Prototype},label={fig:spectre-v2}]
  void benign_function(unsigned long x) {
    return;
  }

  void victim_function(unsigned long x) {
    if (x < array1_size)
      y = array2[array1[x] * 256];
  }

  void dispatcher(func_t func, unsigned long x) {
    (*func)(x);
  }

  int main() {
    /* Attacker: train BTB */
    dispatcher(&victim_function, in_bounds_x);

    /* Victim: run */
    dispatcher(&benign_function, malicious_x);

    /* Attacker: probe cache accesses */
    probe_cache();
  }
\end{lstlisting}
\end{figure}

We first constructed a working proof-of-concept Spectre Variant-2
attack on POWER.  To construct this attack, we
took the original Spectre Variant-1 attack code in
C~\cite{Spectre:Oakland19},
ported it to POWER,
modified it to perform both Spectre Variant-1 and Variant-2
attacks, and reused the cache side channel code that verifies the data
leakage.
Listing~\ref{fig:spectre-v2} shows the core component of our
attack code, abstracting away other technical details to improve
clarity.  In our proof-of-concept attack, the attacker
and the victim share the same virtual address space, which is a common
scenario (e.g., malicious JavaScript code attacking a web browser).
First, the attacker calls {\tt dispatcher} with a
pointer to {\tt victim\_function} and an in-bounds input, mistraining the
processor's branch predictor and BTB to trick the processor into
believing that the function pointer call at Line 11 always branches to
{\tt victim\_function} and that the conditional branch at Line 6 is
always taken.  After training, the victim then calls {\tt dispatcher}
with a pointer to
{\tt benign\_function} and an attacker-supplied malicious input.  Since
the processor has been mistrained, speculative execution will jump
to {\tt victim\_function}, take the conditional branch, and perform
memory accesses that brings contents of {\tt array2} into the
cache.  Even though the processor eventually squashes the speculative reads,
the attacker
infers the secret data via a cache side channel in {\tt probe\_cache}.

The above attack works on POWER when we compile the attack code using
Clang without {\System}.  When we compile the code with
{\System} with fences, the attack no longer works, showing that {\System}
effectively prevents the Spectre Variant-2 attack.

    %%=============================================================================
\section{Space and Performance Evaluation}
\label{section:results}
%%=============================================================================
% need to write introduction for this section

%%-----------------------------------------------------------------------------
\subsection{Experimental Setup}
\label{section:results:setup}
%%-----------------------------------------------------------------------------

We evaluated {\System} by compiling and running the SPEC CPU 2017 benchmark
suite and several real-world applications (Nginx, GnuPG, and
ClamAV) on a
POWER machine.  We used a 64-bit model 2.1 (pvr 004b 0201) 20-core IBM
POWER8 machine running at 4.1~GHz.  The machine, running CentOS 7
Linux with kernel version 3.10.0, has 8 threads per core and 64~GB of RAM.
We compiled the SPEC CPU 2017 benchmarks and applications with both
the original LLVM 4.0.1 (as the baseline) and {\System}.
We statically linked each benchmark program and application; as SFI on
stores requires code and data be within separate virtual address
regions, static linking gives us the easiest control of where code and
data sections are loaded.  In our experiments, we constantly used 32
bytes as the bundle size which was determined experimentally to be the
best choice on POWER ISA.

For SPEC, we used
LLVM's {\it lit} tool to run the SPEC
benchmarks.  It measures both the execution time and code size of the
benchmark programs.  Of all the SPEC benchmark programs that LLVM can
compile (programs written in C or C++ or both), only
{\it 526.blender\_r} does not build on our POWER machine because of an
incompatible C++ header file.  We therefore exclude it from our
experiments.

We begin by evaluating {\System}'s performance
on SPEC when {\System} uses fences to mitigate Spectre attacks.
We then show {\System}'s performance on SPEC
when {\System} uses Spectre-resistant SFI.
Finally, we evaluate {\System}'s overhead on
real-world applications.

%%-----------------------------------------------------------------------------
\subsection{Complete Spectre Defense with Fences}
\label{section:results:eieio}
%%-----------------------------------------------------------------------------

To show how {\System} performs with existing Spectre defenses, we use
{\System} plus SFI on stores and {\tt eieio}
instructions inserted before the first load in each bundle
as a complete system that defends against Spectre Variant-1, Variant-2,
and Spectre variants that poison the RSB.  To study the sources of
overhead, we break down the
overhead into {\System} with just the creation and alignment of
bundles (dubbed {\bf Alignment}), {\System} with the alignment
overhead and the overhead of bit-masking control data
(dubbed {\bf CFI}), {\System} with alignment, control data
bit-masking, and SFI on stores (dubbed {\bf SFI-Store}), and the full
protection with {\tt eieio} instructions (dubbed {\bf Fence}).
We evaluate this system and analyze {\System}'s
impact on the code size and performance of SPEC benchmark programs by
comparing it with the baseline (i.e., the original program compiled
with the same compiler with no {\System} transformations enabled).

%%.............................................................................
\paragraph{Code Size}
\label{section:results:eieio:memory}
%%.............................................................................

\begin{table}[tb]
\centering
\caption{SPEC CPU 2017 Baseline Code Size}
\label{table:spec-code-size-baseline}
\footnotesize{%
\begin{tabular}{|l|r||l|r|}
\hline
  {\bf Benchmark} & {\bf Size (Byte)}
  &
  {\bf Benchmark} & {\bf Size (Byte)}
  \\
\hline
  500.perlbench\_r & 2,747,020
  &
  557.xz\_r & 681,420
  \\
\hline
  502.gcc\_r & 9,579,404
  &
  600.perlbench\_s & 2,747,020
  \\
\hline
  505.mcf\_r & 551,852
  &
  602.gcc\_s & 9,579,404
  \\
\hline
  508.namd\_r & 1,979,680
  &
  605.mcf\_s & 551,852
  \\
\hline
  510.parest\_r & 11,339,744
  &
  619.lbm\_s & 553,900
  \\
\hline
  511.povray\_r & 1,695,168
  &
  620.omnetpp\_s & 3,664,352
  \\
\hline
  519.lbm\_r & 553,804
  &
  623.xalancbmk\_s & 6,124,896
  \\
\hline
  520.omnetpp\_r & 3,664,352
  &
  625.x264\_s & 1,102,188
  \\
\hline
  523.xalancbmk\_r & 6,124,896
  &
  631.deepsjeng\_s & 609,964
  \\
\hline
  525.x264\_r & 1,102,188
  &
  638.imagick\_s & 2,686,828
  \\
\hline
  531.deepsjeng\_r & 609,964
  &
  641.leela\_s & 1,354,144
  \\
\hline
  538.imagick\_r & 2,686,828
  &
  644.nab\_s & 811,660
  \\
\hline
  541.leela\_r & 1,354,144
  &
  657.xz\_s & 681,420
  \\
\hline
  544.nab\_r & 811,660
  &
  &
  \\
\hline
\end{tabular}
}
\end{table}

\begin{figure*}[t]
\centering
\resizebox{\textwidth}{!}{%
  \includegraphics{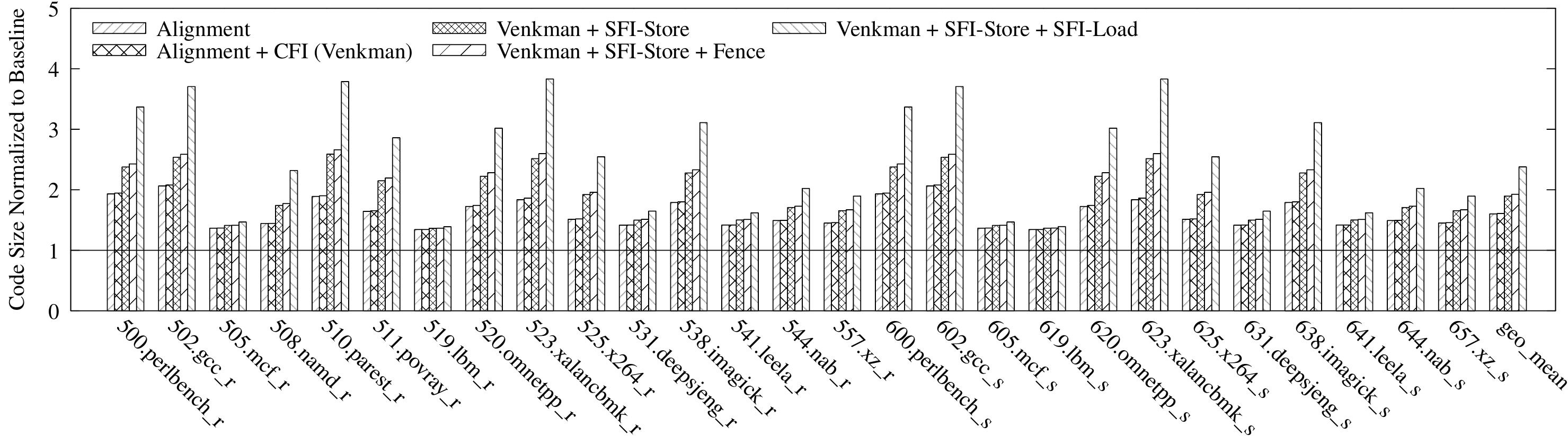}
}
\caption{Code Size Overhead on SPEC CPU 2017}
\label{fig:spec-code-size}
\end{figure*}

We got the code segment sizes of SPEC benchmarks from the {\it lit} tool
running the benchmarks.  The code size information is
actually measured using the {\it llvm-size} tool by reading the ELF
binaries and reporting size of the text segment.  Since we compiled the
benchmark programs statically, the text segment of a program contains
all the library code that the program uses.
Table~\ref{table:spec-code-size-baseline} shows the text segment size of
the SPEC benchmarks compiled by the original Clang, which is our
baseline.  Figure~\ref{fig:spec-code-size} shows the text segment size
of the SPEC benchmarks; the results are normalized to the baseline.
It reports the results for our complete system as well as the breakdown
of the code size overhead, showing how much the overhead is coming from
each of the above defenses.

When all the defenses are deployed, the code size overhead ranges from
1.37$\times$ to 2.66$\times$ with a geometric mean of 1.93$\times$.
As Figure~\ref{fig:spec-code-size} shows, a significant component of the
space overhead comes from Alignment (from
1.34$\times$ to 2.08$\times$ with a geometric mean of 1.61$\times$).
SFI-Store also incurs a non-negligible portion of the overhead
(from 2.16\% to 68.8\% with a geometric mean of 30.1\%) since we
instrumented every store using 3 to 7 instructions.  The rest of the
defenses contributes minor overhead: CFI overhead is from 0.017\% to
2.6\% with a geometric mean of 0.83\%, and Fence overhead is from
0.18\% to 8.3\% with a geometric mean of 3.5\%.

%%.............................................................................
\paragraph{Performance}
\label{section:results:eieio:speed}
%%.............................................................................

\begin{table}[tb]
\centering
\caption{SPEC CPU 2017 Baseline Execution Time}
\label{table:spec-baseline}
\footnotesize{%
\begin{tabular}{|l|r||l|r|}
\hline
  {\bf Benchmark} & {\bf Time (s)}
  &
  {\bf Benchmark} & {\bf Time (s)}
  \\
\hline
  500.perlbench\_r & 48.6
  &
  557.xz\_r & 54.7
  \\
\hline
  502.gcc\_r & 68.4
  &
  600.perlbench\_s & 48.0
  \\
\hline
  505.mcf\_r & 64.0
  &
  602.gcc\_s & 69.3
  \\
\hline
  508.namd\_r & 52.1
  &
  605.mcf\_s & 65.0
  \\
\hline
  510.parest\_r & 68.8
  &
  619.lbm\_s & 237.3
  \\
\hline
  511.povray\_r & 10.1
  &
  620.omnetpp\_s & 103.6
  \\
\hline
  519.lbm\_r & 30.1
  &
  623.xalancbmk\_s & 115.3
  \\
\hline
  520.omnetpp\_r & 99.2
  &
  625.x264\_s & 74.1
  \\
\hline
  523.xalancbmk\_r & 114.9
  &
  631.deepsjeng\_s & 121.1
  \\
\hline
  525.x264\_r & 73.8
  &
  638.imagick\_s & 84.0
  \\
\hline
  531.deepsjeng\_r & 99.6
  &
  641.leela\_s & 132.3
  \\
\hline
  538.imagick\_r & 83.7
  &
  644.nab\_s & 241.7
  \\
\hline
  541.leela\_r & 133.1
  &
  657.xz\_s & 50.2
  \\
\hline
  544.nab\_r & 241.7
  &
  &
  \\
\hline
\end{tabular}
}
\end{table}

\begin{figure*}[t]
\centering
\resizebox{\textwidth}{!}{%
  \includegraphics{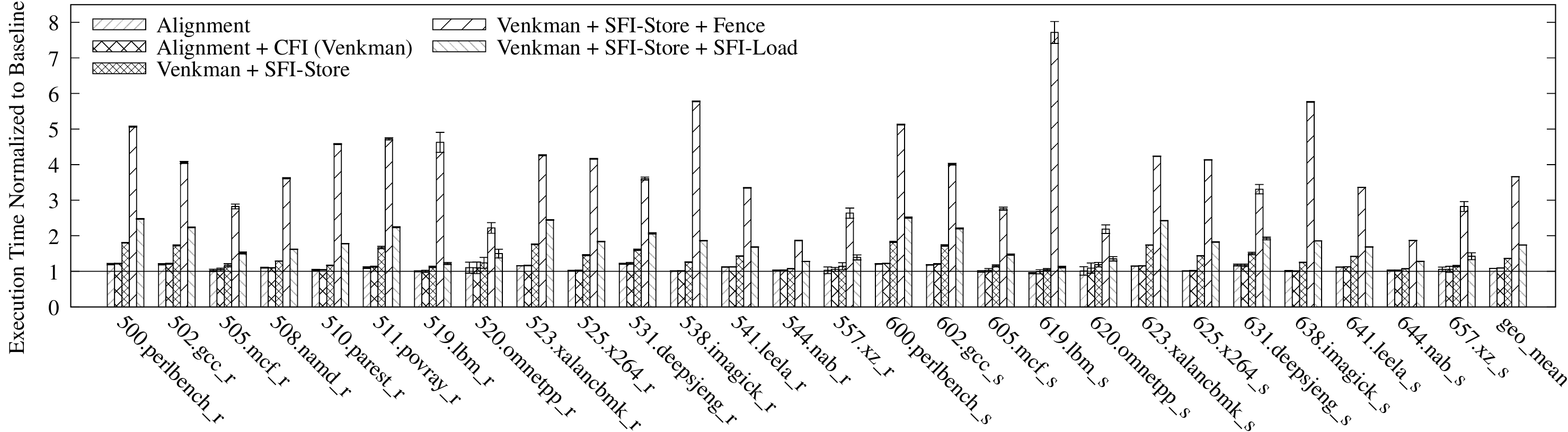}
}
\caption{Performance Overhead on SPEC CPU 2017}
\label{fig:spec-perf}
\end{figure*}

Table~\ref{table:spec-baseline} shows the baseline execution time of the
SPEC benchmarks.  Figure~\ref{fig:spec-perf} shows the
normalized overhead on SPEC benchmarks.  The results include the
overhead of our complete system, while also providing a breakdown of
the performance overhead.  The execution time is measured by the LLVM
Test Suite timing tool called {\it timeit}, which measures the total
wall time of a given command by recording the time difference of two
{\tt gettimeofday()} calls.  We ran
each benchmark ten times and report the geometric mean.

As Figure~\ref{fig:spec-perf} shows, the overhead with all defenses
enabled ranges from 1.87$\times$ to 7.72$\times$ with a
geometric mean of 3.66$\times$.
The performance breakdown shows that, for most of benchmark
programs, Fence and SFI-Store degrade the performance most; the overhead
is from 1.79$\times$ to 7.67$\times$ with a geometric mean of
3.29$\times$ and from 1.04$\times$ to 1.61$\times$ with a geometric mean
of 1.28$\times$,
respectively.  In contrast, the {\System} defenses (i.e., Alignment and
CFI) only have moderate overhead: Alignment incurs overhead from
0.95$\times$ to 1.21$\times$ with a geometric mean of 1.08$\times$,
while CFI incurs overhead from 0.99$\times$ to 1.08$\times$ with a
geometric mean of 1.01$\times$.
%On some benchmark programs such as
%{\it 619.lbm\_s}, {\System} performs even better than the baseline.  We
%investigated such cases and found out that inserting NOPs into these
%programs somehow decreases the number of processor's backend stalled
%cycles and thus improves the performance.

Note that we need SFI-Store to protect the code segment from being
speculatively overwritten only if the processor's store buffer forwards
its content to the instruction fetch unit.  If one needs {\System} to
defend against Spectre attacks and is sure that the processor on which
{\System} is deployed does not forward data in the store buffer to the
instruction fetch unit (which is typically the case if the processor
disallows self-modifying code), then he/she can regain the performance
lost of using SFI-Store by simply disabling SFI-Store.

%%-----------------------------------------------------------------------------
\subsection{Spectre-Resistant Sandboxing}
\label{section:results:sfi}
%%-----------------------------------------------------------------------------

In addition to the evaluation of {\System} with Fence, We also did
experiments on a Spectre-resistant sandboxing system in which Alignment,
CFI, and SFI-Store are still deployed and Fence is replaced with SFI on
loads (dubbed {\bf SFI-Load}), as Section~\ref{section:impl:sfi}
describes.

Figure~\ref{fig:spec-code-size} reports the code size overhead
of the sandboxing SPEC normalized to the same baseline as in
Section~\ref{section:results:eieio}.  As
Figure~\ref{fig:spec-code-size} shows, the overall code size
overhead of the sandboxing ranges from 1.39$\times$ to 3.83$\times$ with
a geometric mean of 2.38$\times$.  SFI-Load's contribution alone is from
1.03$\times$ to 2.32$\times$ with a geometric mean of 1.52$\times$.
Compared with using Fence, the sandboxing approach occupies more storage
space because every load is instrumented with 1 to 3 instructions,
whereas only a single {\tt eieio} instruction is inserted before all the
loads in a bundle using Fence.

Figure~\ref{fig:spec-perf} reports the performance
overhead of the sandboxing normalized to the same baseline as in
Section~\ref{section:results:eieio}.  As
Figure~\ref{fig:spec-perf} shows, the sandboxing
approach slows down the performance by 1.12$\times$ to 2.51$\times$ with
a geometric mean of 1.74$\times$.  Separating the overhead apart,
SFI-Load causes a slowdown of 7.1\% to 68.6\% with a geometric mean of
38.8\%.  Compared with Fence, SFI-Load boosts the performance of
all the benchmarks, confirming the conclusion that using
hardware fences are much more expensive than creating data dependencies
between protecting and protected instructions~\cite{SpectreVG:HASP18}.

%%-----------------------------------------------------------------------------
\subsection{Application Evaluation}
\label{section:results:apps}
%%-----------------------------------------------------------------------------

In addition to evaluating standardized benchmarks like SPEC, we also evaluated
{\System} on selected real-world applications including Nginx, GnuPG,
and ClamAV.  We chose these applications because each one represents a
different type of workload (specifically, I/O intensive, CPU intensive, and
file system intensive).  These programs may also fall victim to
Spectre attacks~\cite{Spectre:Oakland19} as they manipulate sensitive
information.

%%.............................................................................
\paragraph{Nginx}
\label{section:results:apps:nginx}
%%.............................................................................

\begin{table*}[tb]
\centering
\caption{Application Code Size}
\label{table:app-code-size}
\footnotesize{%
\begin{tabularx}{0.9\textwidth}{|r|l|l|X|X|X|X|}
\hline
  {\bf Application} & {\bf Baseline (MB)} & {\bf Alignment} &
  {\bf Alignment + CFI ({\System})} & {\bf {\System} + SFI-Store} &
  {\bf {\System} + SFI-Store + Fence} & {\bf {\System} + SFI-Store + SFI-Load} \\
\hline
  Nginx & 1.20 & 1.61$\times$ & 1.62$\times$ & 2.01$\times$ & 2.04$\times$ & 2.68$\times$ \\
\hline
  GnuPG & 1.91 & 1.62$\times$ & 1.63$\times$ & 2.20$\times$ & 2.27$\times$ & 3.53$\times$ \\
\hline
  ClamAV & 1.14 & 1.71$\times$ & 1.71$\times$ & 2.02$\times$ & 2.05$\times$ & 2.52$\times$ \\
\hline
\end{tabularx}
}
\end{table*}

%\begin{figure}[tb]
%\centering
%\resizebox{\columnwidth}{!}{%
%  \includegraphics{figs/nginx-throughput-baseline}
%}
%\caption{Nginx Baseline Average Throughput}
%\label{fig:nginx-throughput-baseline}
%\end{figure}
\begin{table*}[tb]
\centering
\caption{Application Baseline Performance Results}
\label{table:app-perf-baseline}
\footnotesize{%
\begin{tabular}{|r|r|r|r||r|r|r|r|}
\hline
  {\bf File Size (KB)} & {\bf Nginx (MB/s)} & {\bf GnuPG (ms)} & {\bf ClamAV (ms)} &
  {\bf File Size (KB)} & {\bf Nginx (MB/s)} & {\bf GnuPG (ms)} & {\bf ClamAV (ms)} \\
\hline
  1 & 14.3 & 8.2 & 73.4
  &
  1,024 & 2247.2 & 178.0 & 144.9
  \\
\hline
  2 & 25.7 & 8.3 & 73.4
  &
  2,048 & 2565.1 & 344.8 & 217.4
  \\
\hline
  4 & 48.3 & 8.4 & 73.4
  &
  4,096 & 2701.3 & 684.1 & 145.8
  \\
\hline
  8 & 94.4 & 8.9 & 73.5
  &
  8,192 & 2655.2 & 1357.9 & 651.4
  \\
\hline
  16 & 185.5 & 9.8 & 73.6
  &
  16,384 & 2473.7 & 2717.5 & 550.2
  \\
\hline
  32 & 344.1 & 12.1 & 73.8
  &
  32,768 & 2326.2 & 5414.0 & 1026.9
  \\
\hline
  64 & 620.4 & 17.3 & 74.4
  &
  65,536 & 2267.6 & - & 1980.6
  \\
\hline
  128 & 977.8 & 27.9 & 75.6
  &
  131,072 & 2319.2 & - & 3889.2
  \\
\hline
  256 & 1438.0 & 49.0 & 77.8
  &
  262,144 & 2316.6 & - & 7701.2
  \\
\hline
  512 & 1874.6 & 91.3 & 82.5
  &
  524,288 & 2319.7 & - & 9335.1
  \\
\hline
\end{tabular}
}
\end{table*}

\begin{figure}[tb]
\centering
\resizebox{\columnwidth}{!}{%
  \includegraphics{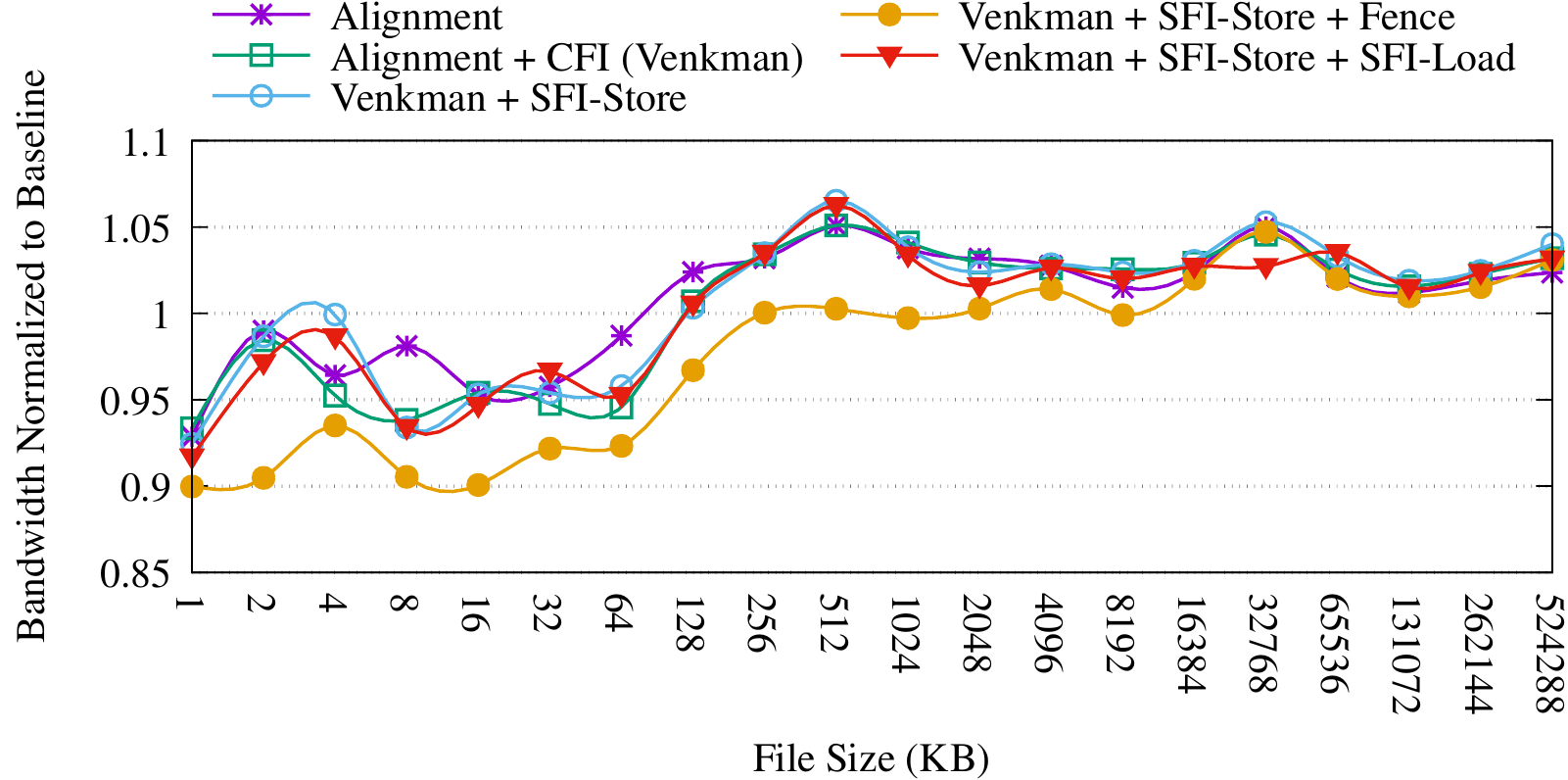}
}
\caption{Throughput Overhead on Nginx}
\label{fig:nginx-throughput}
\end{figure}

We evaluated Nginx 1.15.8 to show how {\System} performs on server
applications.  Nginx~\cite{Nginx} is an open source web server that is
designed for high performance and is widely deployed.
We compiled Nginx with the same configurations as SPEC:
baseline, {\System} with SFI-Store and Fence, {\System} with SFI-Store
and SFI-Load, and configurations that enable us to separate the overhead.

We obtained the code segment size of Nginx using the {\tt objdump} tool.
Table~\ref{table:app-code-size} shows both the baseline code size of
Nginx as well as the code size overhead induced by {\System}.  It also
reports the breakdown of the overhead.
Table~\ref{table:app-code-size} shows that the overall space
overhead is 2.04$\times$ for Fence and 2.68$\times$ for
SFI-Load.  Of all the defenses, SFI-Load, Alignment, and
SFI-Store contribute most of the space overhead, which are 67\%, 61\%,
and 39\%, respectively.  The rest of the defenses add little space overhead:
CFI induces 1\%; Fence induces 3\%.  This roughly conforms with
SPEC's code size overhead shown in
Sections~\ref{section:results:eieio} and~\ref{section:results:sfi}.

To study {\System}'s impact on Nginx performance, we ran Nginx compiled
by {\System} with 1 worker process delivering static files
ranging in size from 1~KB to 512~MB.  The files were
generated by reading bytes from the {\tt /dev/urandom} pseudorandom
number generator.  We used ApacheBench ({\tt ab})~\cite{ApacheBench:2.4} as
the client running on the same machine as Nginx to measure Nginx's
performance.  Since our test machine has 160 logical cores, the client and
server can run on different logical cores without stealing CPU time from
each other when executed on the
same machine.  For each configuration and for each file size,
we ran {\tt ab} for 50
iterations, each iteration lasting 10 seconds in which {\tt ab}
continuously fetched the same file until timeout.  We then collected
performance numbers from the ApacheBench output and report the geometric
means.
Table~\ref{table:app-perf-baseline} shows the baseline average
file transfer throughput of Nginx, and Figure~\ref{fig:nginx-throughput}
shows {\System}'s overhead on Nginx throughput normalized to baseline.
Due to space, we omit numbers on Nginx latency; they are
usually the reciprocal of the throughput.  Standard deviations are also
not shown; they are as much as 4.4\%.
As Figure~\ref{fig:nginx-throughput} shows,
{\System} reduces Nginx's throughput by at most 10.0\% when
using Fence and 8.3\% when using SFI-Load.  {\System} incurs higher
overhead when transferring small files.  As file size increases, the
overhead becomes negligible.  Overall, {\System}'s impact on Nginx
performance is small but not as clear as on SPEC due to high
standard deviations.

%%.............................................................................
\paragraph{GnuPG}
\label{section:results:apps:gnupg}
%%.............................................................................

%\input{tables/gnupg-encrypt-baseline}

\begin{figure}[tb]
\centering
\resizebox{\columnwidth}{!}{%
  \includegraphics{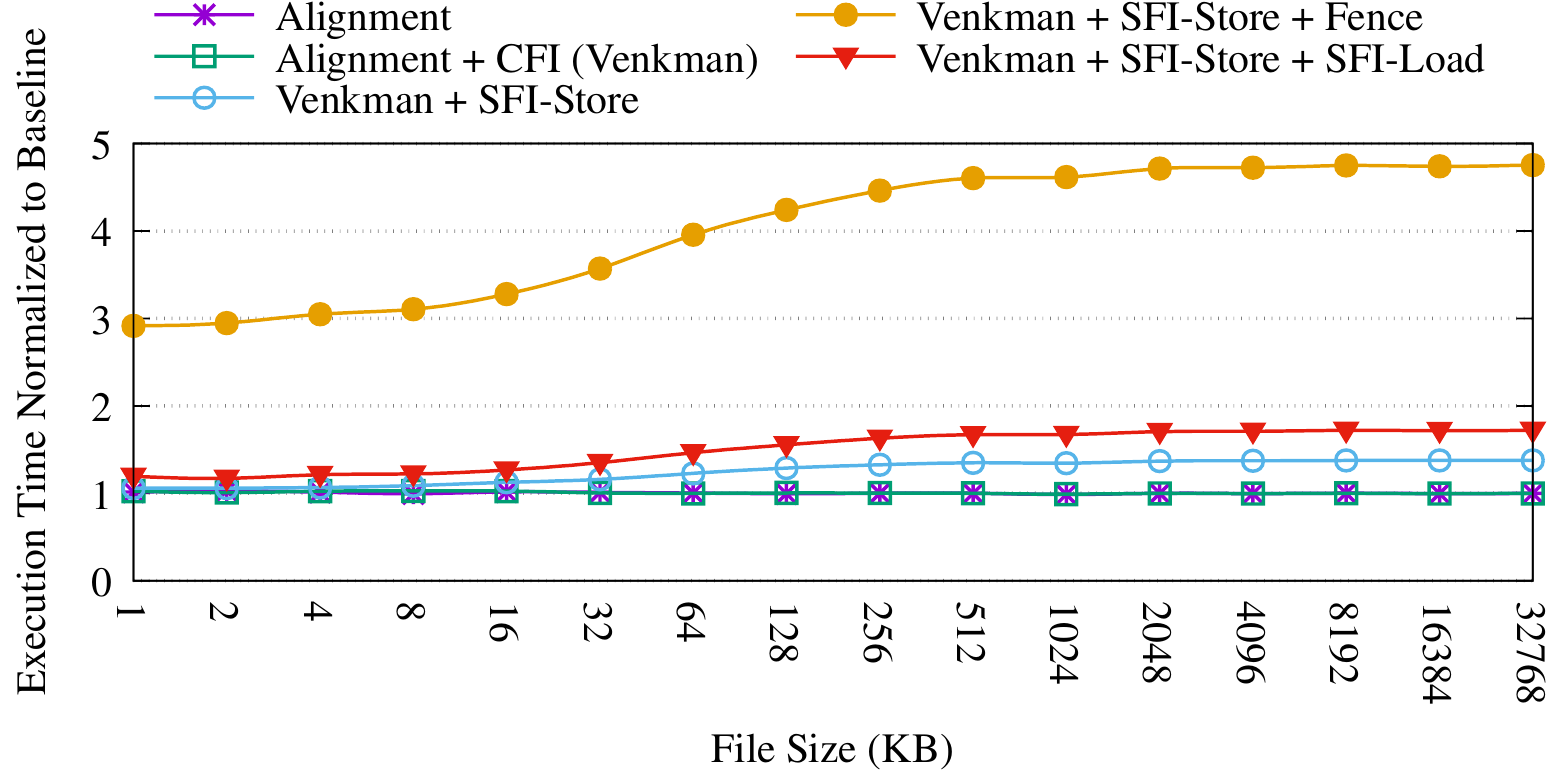}
}
\caption{Performance Overhead on GnuPG Encryption}
\label{fig:gnupg-encrypt}
\end{figure}

We evaluated the code size and performance impact of {\System} on
GnuPG 1.4.23.
GnuPG~\cite{GnuPG} is an open source cryptography program that provides
encryption and signing services.  We compiled GnuPG with the same
configurations as we did for SPEC and Nginx.

Table~\ref{table:app-code-size} provides the code size information of
GnuPG.  This information was also collected by the {\tt objdump} tool.
As Table~\ref{table:app-code-size} shows, when SFI-Load is in place, it
contributes to a significant portion of GnuPG's code size overhead,
resulting in a 2.33$\times$ larger code segment.  Alignment and
SFI-Store also cause a non-negligible increase of GnuPG code size, making
it larger by 62\% and 57\%, respectively.  In contrast, CFI and Fence
incur minor space overhead on GnuPG.

We ran GnuPG compiled by {\System} to encrypt, decrypt, sign, and verify
the signatures of files from 1~KB to 32~MB in size.  The files are a
subset of the files we used in evaluating Nginx.  For each configuration
and for each file size, we ran GnuPG 10 times and recorded the execution
time of each iteration.  The geometric mean over all 10 iterations is
calculated and reported.  Due to limited space, we only show the GnuPG
encryption results which have the highest performance overhead among all
four functionalities we tested; the other three exhibited similar
but lower overhead.  Table~\ref{table:app-perf-baseline}
lists the baseline execution time of GnuPG encryption, and
Figure~\ref{fig:gnupg-encrypt} shows the normalized execution time of
GnuPG compiled by {\System}.  The standard deviations are within
acceptable ranges and thus not shown. As Figure~\ref{fig:gnupg-encrypt} shows,
Fence hurts performance most, increasing the execution time to
2.86$\times$ to 4.38$\times$ with a geometric mean of 3.73$\times$.  If
we adopt the sandboxing approach instead, SFI-Load only gives us a
34.4\% performance degradation at most.  SFI-Store also reports a slowdown
of at most 37.9\%, while Alignment and CFI incur a minor overhead
of at most 2.6\% and 3.1\%, respectively.

%%.............................................................................
\paragraph{ClamAV}
\label{section:results:apps:clamav}
%%.............................................................................

%\input{tables/clamav-baseline}

\begin{figure}[tb]
\centering
\resizebox{\columnwidth}{!}{%
  \includegraphics{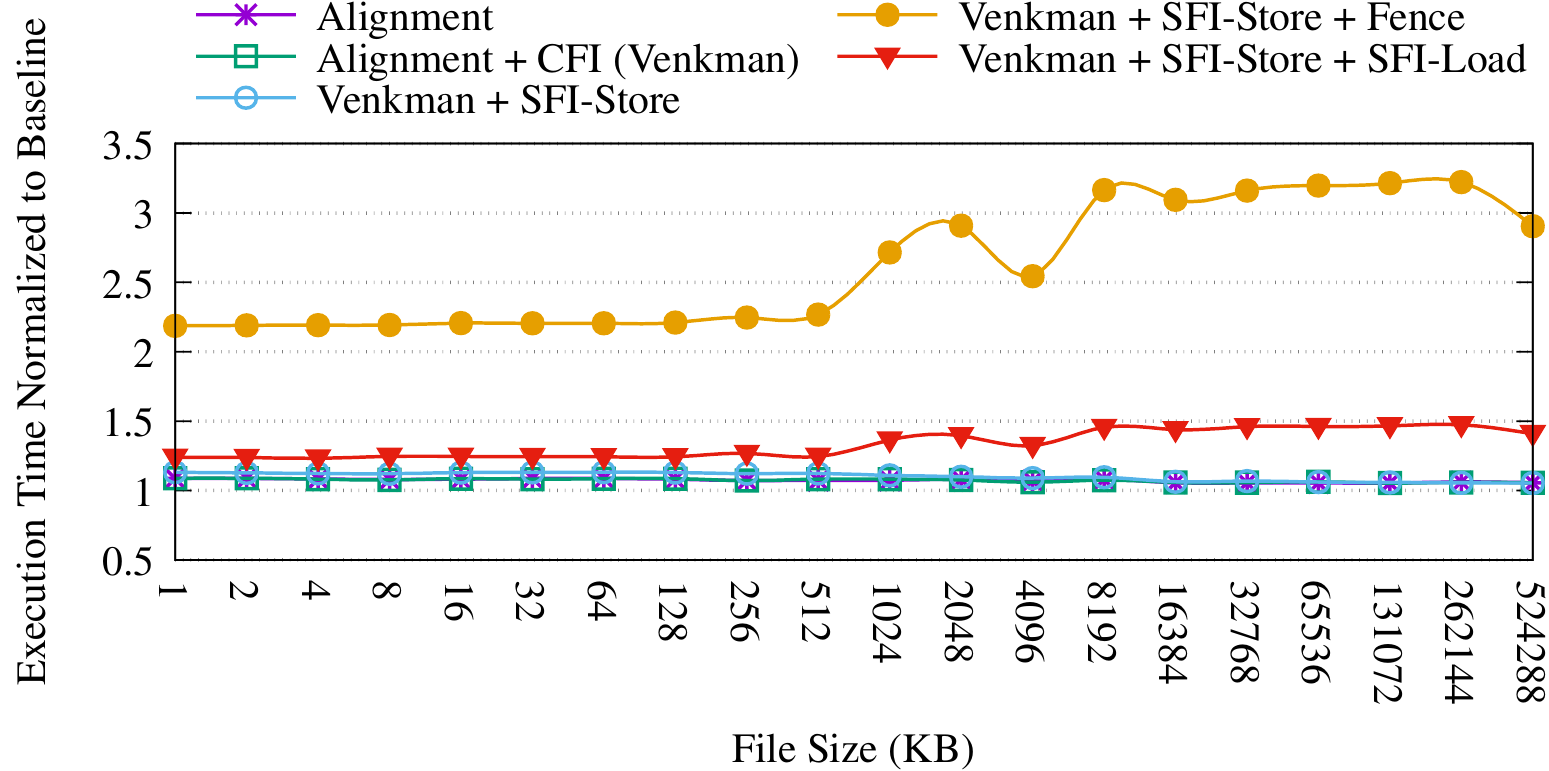}
}
\caption{Performance Overhead on ClamAV}
\label{fig:clamav}
\end{figure}

ClamAV~\cite{ClamAV} is an open source antivirus program that is
typically used for detecting viruses and malware on mail servers.  We
compiled ClamAV 0.92 from the LLVM Test Suite using {\System} with the
same configurations as we did for SPEC, Nginx, and GnuPG, and we report
{\System}'s overhead on ClamAV's command-line scanning tool {\tt clamscan}.

Again, Table~\ref{table:app-code-size} shows the code size measurements
of ClamAV obtained from {\tt objdump}.  When
complete defenses against Spectre attacks are deployed,
Alignment, SFI-Load, and SFI-Store are the three major sources of
ClamAV's code size overhead (71\%, 50\%, and 31\%).  Fence and CFI, on
the other hand, only expand the code segment slightly (3\% and less than
1\%, respectively).

To test ClamAV, we used {\tt clamscan} to scan the files used in the
Nginx experiment for malware.
We ran {\tt clamscan} 10 times for each configuration and
for each file size and report the geometric means.  Our ClamAV uses a
virus database from the LLVM Test Suite; while it is old, it works for
our performance evaluation.  Table~\ref{table:app-perf-baseline}
reports the baseline execution time of ClamAV scanning various-sized
files; Figure~\ref{fig:clamav} shows the performance overhead
incurred by {\System}.  The standard deviations, again, are not shown;
they are acceptable with respect to each configuration.
Figure~\ref{fig:clamav} shows that
Fence reduces performance most, by 2.05$\times$ to 3.17$\times$
with a geometric mean of 2.47$\times$.  In comparison, SFI-Load's
overhead (at most 41.8\%) is much cheaper than Fence.  SFI-Store on ClamAV
causes unusually less overhead (at most 5.0\%), and
{\System} (Alignment and CFI) adds little to minor overhead.

    %%=============================================================================
\section{Related Work}
\label{section:related}
%%=============================================================================

There are several software-only approaches that mitigate Spectre
Variant-1~\cite{Spectre:Oakland19}.
Intel recommends inserting load fences before load instructions
to ensure that branch instructions retire before loads are
executed~\cite{IntelSpecMitigate}.
Carruth proposed pointer hardening~\cite{chandler} which
creates a data dependence between branch conditions
and the pointer used in loads.  The compiler inserts code before loads
that will mask the pointer value to zero if the branch was mispredicted
and leave the pointer unaltered otherwise.
Dong et al.~\cite{SpectreVG:HASP18} developed SFI
techniques that work against Spectre Variant-1 attacks.
All of these approaches are vulnerable to Spectre Variant-2 attacks
since poisoning the BTB~\cite{Spectre:Oakland19} or
RSB~\cite{SpectreRSB:WOOT18,Ret2Spec:CCS18}
permits an attacker to jump over the instructions that protect loads.
{\System} can protect these approaches from Spectre Variant-2 attacks
if it places instructions protecting each load in the
same bundle as the load itself.  {\System} already does this
for load fences~\cite{IntelSpecMitigate} and
SFI~\cite{SpectreVG:HASP18}.  Pointer hardening may require a large
bundle size in order to ensure that the bit-masking instructions and
the branch condition are computed within the same bundle; we leave the
integration of pointer hardening with {\System} to future work.

Spectre Variant-2 poisons the BTB~\cite{Spectre:Oakland19}, and other
variants poison the
RSB~\cite{SpectreRSB:WOOT18,Ret2Spec:CCS18}.  An early mitigation for
Spectre Variant-2, the Retpoline~\cite{Retpoline} transformation,
changes indirect call and jump instructions into return instructions.
A retpoline explicitly moves the target address of an
indirect function call or jump to the return address on the stack,
causing speculation to predict the target address with the RSB instead
of with the BTB.  The
retpoline sets up the RSB so that the processor speculatively executes
a busy loop until the target of the return is read from
the stack.  As Section~\ref{section:spectre:indirect}
explains, the processor uses the BTB to predict the target addresses
of direct branches, making direct branches susceptible to BTB poisoning.
In contrast, {\System} mitigates exploitation of the BTB and RSB
by forcing all control flow targets to be aligned to a bundle's start
address.
%Unlike retpolines, {\System} does not stall speculative execution when
%executing branches.

Intel processors provide three hardware
mitigations for BTB poisoning~\cite{IntelSpecMitigate}.
The first prevents code running at a lower privilege level from affecting
branch prediction for code running at higher privilege levels.
Unfortunately, several Spectre attacks~\cite{Spectre:Oakland19} target
victims running at the same hardware privilege level as the attacker.  In
contrast, {\System} protects software running at all hardware privileges levels.
The second Intel processor defense prevents sharing of BTB entries
between code running on different logical processors.  This defense
fails to mitigate attacks by programs executing on the same logical
processor as the victim.  {\System}, on the other hand, protects
software regardless of which logical processors execute the attacker
and victim code.
The third defense adds a BTB training barrier command: branch
predictions following the barrier do not use BTB entries that were
created on the same logical processor prior to the
barrier.  This approach prevents BTB
poisoning but reduces performance as valid BTB entries prior to the
barrier are lost.  {\System} ensures that all BTB and RSB entries are
properly aligned on a bundle boundary, allowing safe sharing of BTB
entries across programs.  Only programs using fences or SFI
instructions incur significant performance loss.

{\System} employs transformations similar to those of
Google's Portable Native Client (PNaCl)~\cite{PNaCL:UsenixSec10}.
Like PNaCl, {\System} must break code into individual bundles,
align bundles on a constant alignment, and align control data (such as
function pointers) before using them in a branch.
However, since {\System} regulates control flow for
\emph{speculatively} executed instructions, it must also place call
instructions at the end of bundles to ensure that return addresses
are aligned to the start of a bundle.  It must also ensure that
protecting instructions (e.g., fences and SFI) and protected
instructions (e.g., loads and stores) be located within the same bundle.

    %%=============================================================================
\section{Conclusions and Future Work}
\label{section:conc}
%%=============================================================================

In this paper, we presented and evaluated {\System},
a solution that thwarts Spectre attacks that poison the BTB
and RSB.  To the best of our knowledge, no existing defense
completely mitigates poisoning of these structures.  Our evaluation
shows that {\System} increases code size by 1.94$\times$ on
average. We also observe
an average of 3.47$\times$ performance overhead.

%
% JTC: I don't want to include the text below unless we have a
% quanitative comparison.
%
%which is negligible
%when compared to the overheads incurred by solutions that disable
%speculative execution.

Several directions exist for future work.  First, we will
improve {\System}'s precision.  {\System} currently ensures that
speculative execution adheres to a very conservative
CFG.  More sophisticated code placement strategies might allow
{\System} to restrict speculative branches to a set of specific targets.
Second, we will investigate whether transformations like
super-block construction~\cite{AllenKennedy}
and
if-conversion~\cite{AllenKennedy}
can reduce the number of fences that {\System} inserts by
creating more straight-line code.
Finally, we will port {\System} to x86 and ARM.

%-------------------------------------------------------------------------------
\bibliographystyle{plain}
\bibliography{venkman,sva,security,safecode,basic,optimization}

%%%%%%%%%%%%%%%%%%%%%%%%%%%%%%%%%%%%%%%%%%%%%%%%%%%%%%%%%%%%%%%%%%%%%%%%%%%%%%%%
\end{document}